\documentclass[final,3p]{elsarticle}

\usepackage{xcolor}         

\journal{}
\date{}
\title{Advancing Automatic Photovoltaic Defect Detection using Semi-Supervised Semantic Segmentation of Electroluminescence Images}

\author[inst1]{Abhishek Jha}

\affiliation[inst1] {
            addressline={Delhi Technological University}, 
            city={New Delhi}, 
            state={Delhi},
            country={India}}

\author[inst2]{Yogesh Rawat}
\author[inst2]{Shruti Vyas*}
\cortext[cor1]{Corresponding author: shruti@ucf.edu}
\affiliation[inst2]{
            addressline={University of Central Florida}, 
            city={Orlando},
            state={Florida},
            country={United States of America}}

\usepackage{amssymb}

\usepackage[utf8]{inputenc} 
\usepackage[T1]{fontenc}    
\usepackage{hyperref}       
\usepackage{url}            
\usepackage{booktabs}       
\usepackage{amsfonts}       
\usepackage{nicefrac}       
\usepackage{microtype}      
\usepackage{xcolor}         
\usepackage{algorithm}
\usepackage{graphicx}
\usepackage{amsmath}
\usepackage{algorithmic}
\usepackage{multicol,multirow}
\usepackage{geometry}
\usepackage{booktabs}
\usepackage{threeparttable}
\usepackage{xspace}
\usepackage{colortbl}
\usepackage{caption}
\usepackage{array}
\usepackage{siunitx}
\usepackage{makecell}
\usepackage{tabularx}
\usepackage[english]{babel}
\usepackage[autostyle]{csquotes}
\usepackage{xcolor}
\usepackage{tcolorbox}
\usepackage{float}
\begin{document}
\begin{frontmatter}
\begin{abstract}

Photovoltaic (PV) systems allow us to tap into all abundant solar energy, however they require regular maintenance for high efficiency and to prevent degradation. Traditional manual health check, using Electroluminescence (EL) imaging, is expensive and logistically challenging which makes automated defect detection essential. 
Current automation approaches require extensive manual expert labeling, which is time-consuming, expensive, and prone to errors. We propose PV-S3 (\textbf{P}hoto\textbf{v}oltaic-\textbf{S}emi-supervised \textbf{S}emantic \textbf{S}egmentation), a Semi-Supervised Learning approach for semantic segmentation of defects in EL images that reduces reliance on extensive labeling. PV-S3 is an artificial intelligence (AI) model trained using a few labeled images along with numerous unlabeled images. We introduce a novel Semi Cross-Entropy loss function to deal with class imbalance. 
We evaluate PV-S3 on multiple datasets and demonstrate its effectiveness and adaptability. 
With merely 20\% labeled samples, we achieve an absolute improvement of 9.7\% in mean Intersection-over-Union (mIoU), 13.5\% in Precision, 29.15\% in Recall, and 20.42\% in F1-Score over prior state-of-the-art supervised method (which uses 100\% labeled samples) on University of Central Florida-Electroluminescence (UCF-EL) dataset (largest dataset available for semantic segmentation of EL images) showing improvement in performance while reducing the annotation costs by 80\%. For more details, visit our GitHub repository: \href{https://github.com/abj247/PV-S3}{https://github.com/abj247/PV-S3}.


\end{abstract}
\end{frontmatter}

\section{Introduction}
\label{sec: introduction}

The increasing global demand for clean and sustainable energy sources has highlighted the importance of solar energy, particularly photovoltaic (PV) systems, in the renewable energy landscape \cite{8645580,kannan2016solar,SHAHSAVARI2018275}. While PV modules are essential for converting sunlight into electricity, their efficiency and reliability are often compromised by various defects, ranging from manufacturing imperfections to environmental stressors \cite{grimaccia2017pv,li2021application,fan2017study,koehl2012modelling}. 
These defects reduce the overall efficiency and the life cycle while impacting the financial viability of PV systems.
Various imaging techniques such as Infrared Thermography (IRT), Photoluminescence (PL) Imaging, and Electroluminescence (EL) imaging  are currently used, however, EL imaging remains a key technique due to its high resolution and detailed electrical property information \cite{trupke2012photoluminescence, kellil2023fault,bauer2009lock, davis2017electroluminescence, vorster2007high}. 
Traditionally, defect detection in PV modules has relied on manual inspection of EL images \cite{bliss2015spatially,dhimish2018pv,dhimish2017impact}. 
Manual inspection however requires certain expertise while being labor-intensive and prone to human error. Recent expansion of PV installations increased the number of images that need to be analyzed thus exacerbating the problem. In long run automation is the key to address all these issues. 

Application of artificial intelligence (AI) to bring automation has been widely accepted in the energy industry \cite{connolly2010review, hwang2023intelligent} and owing to above challenges, application for automated defect detection in PV modules has also gained attention \cite{abubakar2021review, ahan2021ai, akram2022failures, sundaram2021deep}. AI-based approaches, including binary classification for defective/non-defective, multi-class classification for identifying types of defects, and object detection for locating defects, pave the way for comprehensive analysis of PV systems \cite{hassan2023enhancing, zhang2020detection}. Among these, semantic segmentation stands out for its ability to delineate and classify each pixel of an image according to defect type, offering a detailed and nuanced understanding of defects. However, despite its potential for enhancing defect detection accuracy and efficiency, semantic segmentation is often constrained by the high costs associated with extensive manual annotation of data \cite{pratt2021defect, fioresi2021automated}. A major obstacle is the requirement for extensive labeled training data which requires expert human labor. 

To overcome this limitation, we propose PV-S3 (\textbf{P}hoto\textbf{v}oltaic-\textbf{S}emi Supervised \textbf{S}emantic \textbf{S}egmentation), a semi-supervised deep learning framework for semantic segmentation in PV module defect detection, which efficiently utilize both labeled and unlabeled EL images. PV-S3 reduces the reliance on extensive labeled data while addressing the scalability issues in large-scale solar installations.  It is based on mean-teacher approach \cite{tarvainen2017mean} and efficiently leverages unlabeled data. The proposed method enhances model accuracy and generalization by enforcing consistency between the predictions of a student model and a temporally averaged teacher model. Such an approach is crucial for reducing reliance on extensive labeled datasets, addressing a key challenge in the semantic segmentation of PV module defects. To the best of our knowledge there is no current existing work in Photovoltaics making use of semi-supervised learning approaches or reduced labeled data for defect detection.

\begin{figure}[h!]
\centering
\includegraphics[width=1.0\textwidth]{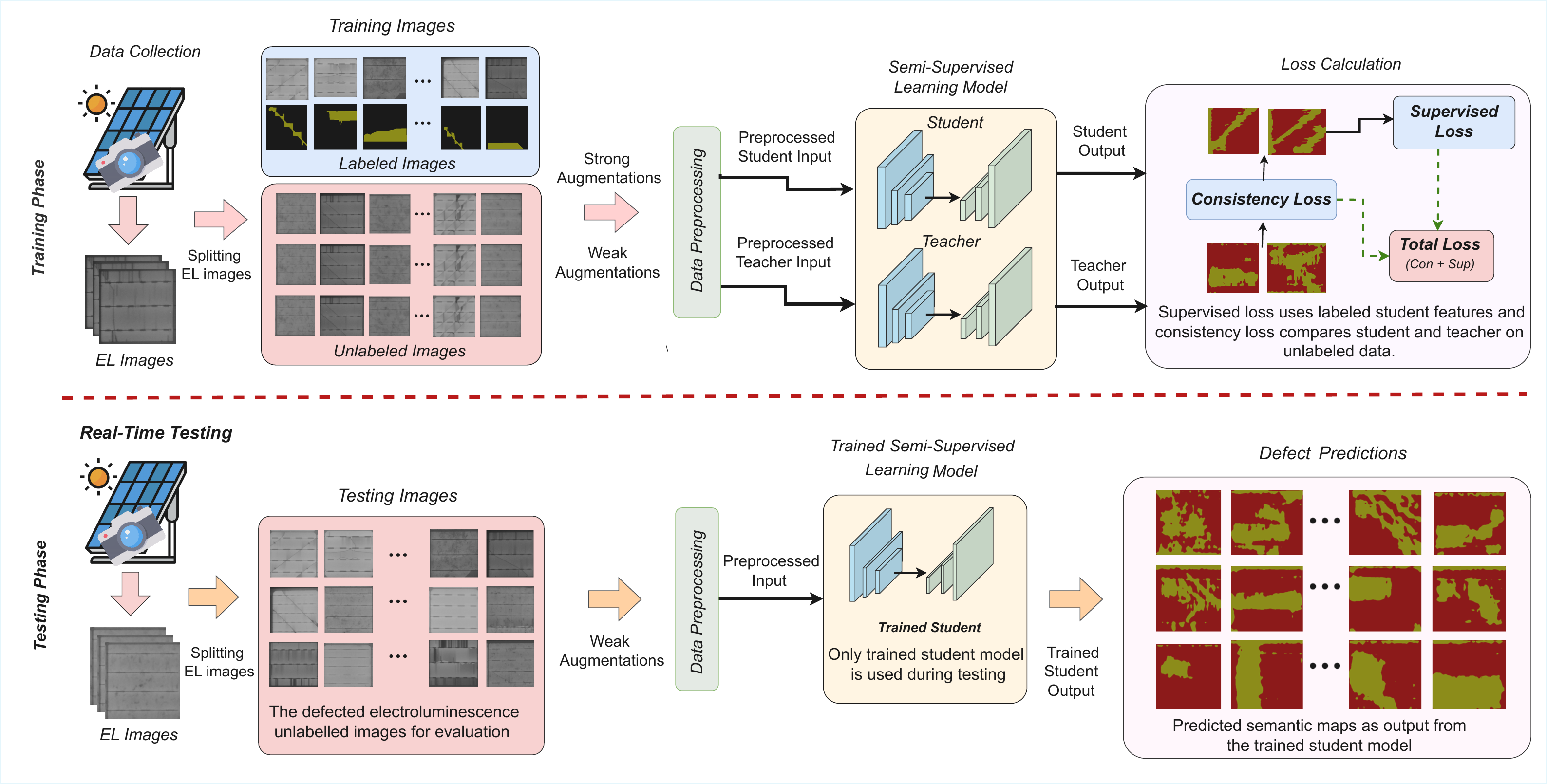}
\caption{\textbf{An overview of the proposed semi-supervised approach for PV defect detection:} It utilizes both labeled and unlabelled images in a semi-supervised learning (SSL) framework  using supervised and unsupervised consistency losses to train the defect detection model (top row) and during testing the trained model predicts the defect class for each pixel in EL images (bottom row).} 
\label{fig: teaser}
\end{figure}

To address the challenges posed by the scarcity of labeled data in real-world datasets, we propose a semi-supervised approach.  We effectively leverages both labeled and unlabeled data to enhance the model's ability to detect defects. Our Semi-cross entropy loss introduces stricter positive and negative thresholds, fostering stronger convergence by deferentially weighting the contributions of various classes. This ensures that the model extracts valuable patterns from the unlabeled data. Figure~\ref{fig: teaser} represents an high-level overview of the proposed framework for defect detection in photovoltaic cells using semi-supervised segmentation.


The major contributions of our research are as follows:
\begin{itemize}
     \item  We study the novel problem of automatic defect detection in photovoltaic (PV) modules under limited annotations, reducing the dependency on large labeled datasets and alleviating the need for extensive manual annotation—a problem not previously explored in existing literature.

    \item We propose PV-S3, a semi-supervised framework for semantic segmentation of PV modules, which effectively leverages both labeled and unlabeled data to enhance defect detection accuracy while maintaining label efficiency.

    \item  We perform extensive evaluation of the proposed approach on several real-world PV datasets, demonstrating its effectiveness and efficiency in detecting various types of defects.
    \item We demonstrate that the proposed approach outperforms existing state-of-the-art fully supervised methods with merely 20\% annotations. 
\end{itemize}


In this study, we employ semi-supervised learning (SSL) for automatic detection of defects in PV modules to reduce the need for extensive annotation. The proposed method will advance the field of PV maintenance by providing an efficient, and cost-effective solution for defect detection.




\section{Related Work}
\label{sec: background}
Defect detection in PV modules has witnessed a progression from manual inspection to advanced automated techniques. Traditional methods involving manual visual examination of EL images have limitations in scalability and subjectivity \cite{gallardo2020nondestructive}. Adoption of machine learning techniques, particularly deep learning with Convolutional Neural Networks (CNNs), has enabled more accurate and efficient detection.
\paragraph{{EL Imaging and PV Cells}}

PV modules are essential components for converting sunlight into electrical energy, comprised of interconnected photovoltaic cells that absorb sunlight to generate direct current (DC) electricity. However, defects introduced during manufacturing, handling, and installation processes can impair their performance and reliability. Common issues include cracks, fractures, soldering problems, corrosion, and electrical contact issues \cite{rajput2018quantitative, tang2024module}. These defects can lead to decreased power output, hotspots, accelerated degradation, and safety risks. Timely detection and identification of these issues are crucial for ensuring optimal PV system operation and longevity.
Two primary techniques used for assessing PV module performance are current-voltage (I-V) measurements and electroluminescence (EL) imaging. EL imaging is particularly valuable for maintenance evaluations and defect detection. Given the large scale of PV plants, there is a growing focus on automating defect detection and classification in EL images \cite{tsai2012defect, meng2022defect}.

\paragraph{Classification}

Defect classification in PV modules is crucial for panel efficiency and longevity. Traditional machine learning methods like Support Vector Machines (SVMs) \cite{9065065}, Random Forests \cite{9065065, chen2018random}, and k-Nearest Neighbors (k-NN) \cite{9065065} along with ensemble learning techniques \cite{sepulveda2024ensemble} have been widely applied, particularly effective for simpler defect types. However, the diverse and complex nature of PV defects demands more advanced techniques. Deep learning, especially Convolutional Neural Networks (CNNs), has gained prominence for its ability to learn from large datasets, capturing subtle defect variations \cite{mellit2022embedded}. Modified CNN variants like multi scale CNNs with transfer learning \cite{korkmaz2022efficient} and multi-scale residual CNNs with class balancing strategy \cite{wu2024feature} have shown effectiveness in defect detection. Recently vision transformer based architectures have also been used for defect detection if PV Cells \cite{dwivedi2024identification}. YOLO based architecture have also been used to detect defects in the electroluminescence images for PV defect detection. YOLOv8 improved architecture is used to get enhanced performance \cite{cao2024improved}. Adaptive fusion integrated with YOLO architectures to expand the backbone feature's receptive field and adaptively filter out conflicting information across varied hierarchical levels improved the detection performance \cite{yang2024novel}. 

\paragraph{Segmentation}

Defect detection, unlike classification, demands defect localization in images, employing segmentation techniques for precise delineation. This task is vital for understanding defect characteristics and their spatial distribution in detail. Numerous models and methodologies aim for accurate segmentation of PV cell defects \cite{fioresi2021automated, pratt2023benchmark}. Semantic segmentation, a robust computer vision technique, has recently gained traction for this purpose, leveraging powerful pre-trained models \cite{fioresi2021automated, pratt2023benchmark}. However, training segmentation models, especially for specialized domains like PV module defect detection, poses unique challenges, primarily due to extensive data annotation requirements \cite{ouali2020overview}. The effectiveness of segmentation models hinges on ample labeled data availability for training, underscoring the necessity for efficient annotation strategies. 

\paragraph{Semi-supervised Segmentation}

Recently, semi-supervised learning (SSL) techniques~\cite{tarvainen2017mean, sohn2020fixmatch} have demonstrated strong potential in segmentation tasks by leveraging both limited labeled and abundant unlabeled data. Approaches such as pseudo-labeling, consistency regularization, and teacher-student frameworks have been effectively applied in domains like medical imaging and industrial defect inspection~\cite{ouali2020overview, yu2019uncertainty, bortsova2019semi, liang2022semi}, enabling improved performance with minimal supervision.
Motivated by these successes, we investigate a semi-supervised segmentation framework tailored for photovoltaic (PV) defect detection using electroluminescence (EL) imagery. To the best of our knowledge, this is the first work on SSL-based segmentation in the context of EL-based PV module inspection. By integrating semantic segmentation with a teacher-student framework, our approach enables effective learning from both labeled and unlabeled data, reducing the reliance on extensive annotations while improving defect localization performance.



\section{Proposed Methodology}
\label{sec: Methodology}

\begin{figure}[h!]
\centering
\includegraphics[width=1.0\textwidth]{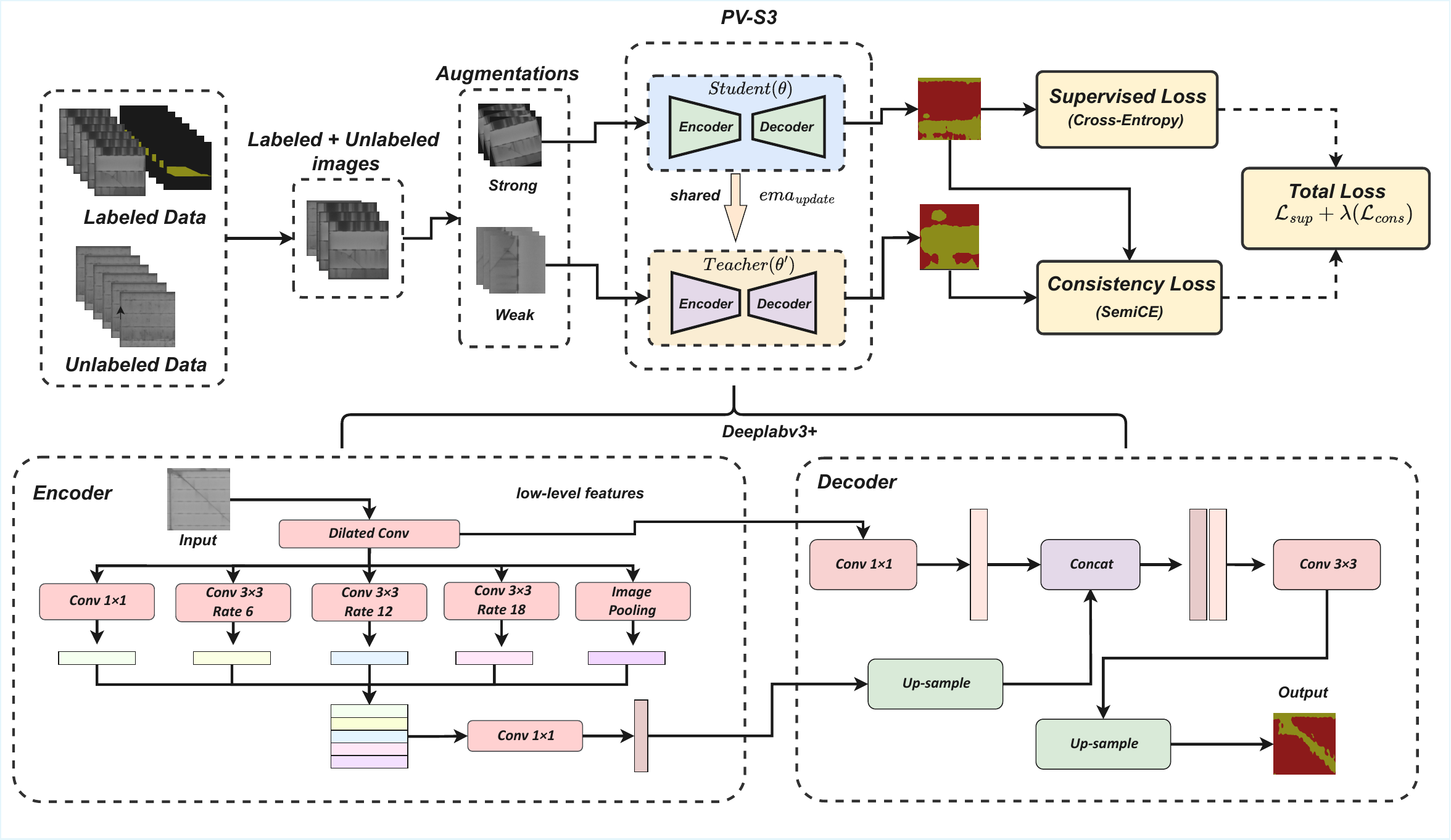}
\caption{\textbf{Overview of PV-S3 for Semantic Segmentation of PV Defects:} The framework takes labeled and unlabeled images as input and both labeled and unlabeled images are given as input to both student and teacher models. The images are augmented before passing to both models where strong augmented images are given to the student network and weak augmented images to the teacher network. Deeplabv3+ is used as segmentation model for both student and teacher networks.
}
\label{fig: Mean Teacher Framework}
\end{figure}

The proposed method addresses the challenge of limited labeled data in PV defect detection through semi-supervised learning. We utilize the mean-teacher approach \cite{tarvainen2017mean}, aiming to enhance defect detection accuracy and efficiency while reducing the need for extensive expert annotation. An overview of the proposed approach is shown in Figure ~\ref{fig: Mean Teacher Framework} displaying the process of PV-S3 for defect detection in photovoltaic cells. 

Given a dataset \(D\) comprising labeled images \(X_{L}\) with corresponding ground truth segmentation masks \(Y_{L}\), and unlabeled images \(X_{U}\) without associated labels, our objective is to train a model \(M\) capable of accurately generating segmentation masks for new, unseen input images. \(X_{L} = \{X_{L}^1, X_{L}^2, ..., X_{L}^{N}\}\) represents the set of \(N\) labeled images, with \(Y_{L} = \{Y_{L}^1, Y_{L}^2, ..., Y_{L}^{N}\}\) denoting the corresponding segmentation masks indicating defect present and type. \(X_{U} = \{X_{U}^1, X_{U}^2, ..., X_{U}^{K}\}\) denotes unlabeled images, where \(K\) is their count. These images lack ground truth masks but are utilized by the model to learn from unlabeled data through semi-supervised techniques. The primary output comprises predicted segmentation masks \(\hat{Y}\) for input images \(X\), with each mask \(\hat{Y}^i\) corresponding to the model's prediction for \(X^i\). Model \(M\) aims to output pixel-wise classifications indicating the likelihood of each pixel belonging to various defect types.




\subsection{Background}
\label{subsec: Model selection}

Semi-supervised learning utilizes both labeled and unlabeled data thus filling gaps in conventional supervised learning. It leverages abundant unlabeled data to enhance model's performance. The Mean Teacher framework \cite{tarvainen2017mean} is the key here, employing a student network \(S(\theta)\) and a teacher network \(T(\theta')\). The teacher network's weights are updated based on the student network through exponential moving average, facilitating pseudo-labeling for unlabeled data. This integration improves learning outcomes by guiding the student to match teacher predictions on labeled data and learn from them on unlabeled data, enhancing model generalization. This approach is especially valuable with limited labeled data, allowing learning from both labeled samples and larger unlabeled datasets. For segmentation tasks, the Mean Teacher framework extends to pixel-wise labeling \cite{liu2022perturbed}, with the teacher's pixel-wise predictions serving as soft labels for unlabeled images. This guidance aids the student in learning the spatial layout of classes, improving segmentation accuracy.



\subsection{Proposed Approach}

PV-S3 is a semi-supervised learning framework leveraging both labeled and unlabeled data to enhance defect detection. It employs a dual-network system: a student network \(S(\theta)\) and a teacher network \(T(\theta')\). The teacher network guides the student using pseudo-labels generated from its predictions on unlabeled data \(X_U\), enabling the student network to learn complex defect patterns without explicit annotations. PV-S3 utilizes two loss functions: cross-entropy for labeled data and Semi Cross-Entropy (SemiCE) loss for unlabeled data, enhancing predictive consistency. 
PV-S3 adopts the DeepLabv3+ architecture with pre-trained ResNet50 weights, featuring an encoder-decoder architecture with atrous convolutions for multi-scale contextual information extraction. The model is trained with the mean teacher approach, updating student parameters via gradient descent and refining teacher parameters with Exponential Moving Average (EMA). This ensures stable guidance and improved generalization.

The algorithm ~\ref{alg:semi_supervised_learning} outlines PV-S3's semi-supervised defect detection in solar PV modules. It begins with labeled data $(X_L, Y_L)$, unlabeled data $X_U$, and hyperparameters such as $\lambda$, $\alpha$, and epochs, producing the trained student model $S(\theta')$. Student and teacher networks $S(\theta)$ and $T(\theta')$ are initialized with same pretrained weights. For each epoch, mini-batches $M_L$ and $M_U$ are sampled from labeled and unlabeled datasets to get training samples ($M_L$, $M_U$). Strong and weak augmentations are applied to these training samples to get ($M_L''$, $M_U''$) and ($M_L'$, $M_U'$).
($M_L''$, $M_U''$) is passed to the Student network and ($M_L'$, $M_U'$) is passed to the Teacher network.
Losses $\mathcal{L}_S$ and $\mathcal{L}_U$ for labeled and unlabeled data are computed using Cross-Entropy and SemiCE functions. Total loss $\mathcal{L}_{total}$, a weighted sum, balances both losses with $\lambda$. Student parameters update via backpropagation based on total loss, while teacher parameters refine using EMA with decay rate $\alpha$.

\begin{algorithm}
\caption{PV-S3: Semi-Supervised Learning for PV Module Defect Detection}
\label{alg:semi_supervised_learning}
\begin{algorithmic}[1]
    \STATE $Input: (X_L, Y_L), X_U$ \hfill \COMMENT{Labeled and unlabeled images}
     \STATE $Hyperparameters: \lambda, \alpha,\text{epochs}$ \hfill   \COMMENT{Hyperparameters for training}
    \STATE $Output: $ \(S(\theta')\) \hfill        \COMMENT{Trained Student network}
    \STATE Initialize $S(\theta), T(\theta')$ \hfill \COMMENT{Initialize networks}
    \FOR{$epoch = 1$ to $\text{epochs}$}
        \FOR{each $batch$ in training batches}
            \STATE $M_L \gets$ sample from $(X_L, Y_L)$ \hfill \COMMENT{Sample labeled mini-batch}
            \STATE $M_U \gets$ sample from $X_U$ \hfill \COMMENT{Sample unlabeled mini-batch}
            \STATE $(M_L',M_U') \gets$ apply weak augmentation to $(M_L,M_U)$ \hfill \COMMENT{Augment mini-batch}
            \STATE $(M_L'',M_U'') \gets$ apply strong augmentation to $(M_L,M_U)$ \hfill \COMMENT{Augment mini-batch}
            \STATE $(P_SL,P_SU) \gets S(M_L'',M_U'')$ \hfill \COMMENT{Compute student predictions on $(M_L'',M_U'')$}
            \STATE $(P_TL,P_TU) \gets T(M_L',M_U')$ \hfill \COMMENT{Compute teacher predictions on $(M_L',M_U')$}
            \STATE $\mathcal{L}_S \gets \text{CrossEntropy}((P_SL, Y_L)$ \hfill \COMMENT{Supervised loss for labeled data}
            \STATE $\mathcal{L}_U \gets \text{SemiCrossEntropy}(<P_SU,P_SL>, <P_TU,P_TU>)$ \hfill \COMMENT{Unsupervised loss}
            \STATE $\mathcal{L} \gets \mathcal{L}_S + \lambda \mathcal{L}_U$ \hfill \COMMENT{Total loss}
            \STATE Update $S$ using $\nabla L$ \hfill \COMMENT{Update student network}
            \STATE $\theta' \gets \alpha\theta' + (1-\alpha)\theta$ \hfill \COMMENT{Update teacher with EMA}  
        \ENDFOR
    \ENDFOR
\end{algorithmic}
\end{algorithm}

\subsubsection{Supervised Loss}
\label{subsubsec: SL}
In the context of the mean teacher network, crucial for defect detection in EL images, the optimization of the supervised loss, along with the consistency loss, is emphasized. The supervised loss is vital, aiming to reduce the difference between the student network's pixel-wise predictions and the actual segmentation masks, which are the ground truth for the labeled dataset.

The supervised loss for this semantic segmentation task is computed using the cross-entropy loss function.  which involves making predictions at the pixel level across multiple classes or defect types. The supervised loss is formulated as follows:
\begin{equation}
\mathcal{L}_{supervised} = -\frac{1}{N_{L}} \sum_{i=1}^{N_{L}} \sum_{c=1}^{C} \sum_{p=1}^{P} Y_{labeled}^{i,p}(c) \log(P_{student}^{i,p}(c))
\end{equation}
Here, \(N_{L}\) represents the total number of labeled images in the dataset, \(C\) denotes the number of defect types, and \(P\) is the total number of pixels in each image. The term \(Y_{labeled}^{i,p}(c)\) indicates whether class \(c\) is present at pixel \(p\) for the \(i\)-th instance, and \(P_{student}^{i,p}(c)\) reflects the student model's predicted probability that pixel \(p\) in instance \(i\) belongs to class \(c\). 

\subsubsection{SemiCE as Consistency Loss}
\label{subsubsec: CL}

The consistency loss ensures uniformity between student and teacher network predictions by comparing their outputs on labeled and unlabeled data. We use SemiCE Loss as consistency loss adapted for semi-supervised defect detection. SemiCE enhances learning in scenarios of data scarcity and class imbalance, focusing on differential treatment of positive and negative predictions across labeled and unlabeled data. The SemiCE loss consists of two key components: the positive loss ($\mathcal{L}_{\text{pos}}$) and the negative loss ($\mathcal{L}_{\text{neg}}$). $\mathcal{L}_{\text{pos}}$ emphasizes improving the accuracy of positive predictions crucial for defect identification. Conversely, $\mathcal{L}_{\text{neg}}$ handles negative predictions below the confidence threshold to prevent excessive noise interference. SemiCE loss enhances training stability and effectiveness by specifically addressing the challenges posed by class imbalance. By optimizing positive and negative predictions while considering unlabeled data characteristics, SemiCE aims on  accurate defect detection in PV modules.

The $\mathcal{L}_{\text{pos}}$ represents the loss associated with confident positive predictions. These are predictions where the model assigns a high probability to the correct class and the input's confidence exceeds a certain threshold ($t_\text{pos}$). In mathematical terms it is formulated as :

\begin{equation}
\mathcal{L}_{\text{pos}} = - \frac{1}{N_{\text{U}}} \sum_{i=1}^{N_{\text{U}}} \sum_{j=1}^{H \times W} \begin{cases}
     Y_{\text{L},ij} \cdot \log \left( \frac{\exp(P_{\text{$\theta$},ij})}{\sum_{k=1}^{H \times W} \exp(P_{\text{$\theta$},ik})} \right), & \text{if } Y_{\text{L},ij} = 1 \text{ and } P_{\text{$\theta$},ij} \geq t_{\text{pos}} \\
    0, & \text{otherwise}.
\end{cases}
\tag{2}
\end{equation}

On the other hand, $\mathcal{L}_{\text{neg}}$ accounts for negative predictions below the confidence threshold. This threshold is set as $t_\text{neg}$ and is typically chosen to ensure that only highly confident negative predictions contribute to the loss. In mathematical terms:

\begin{equation}
\mathcal{L}_{\text{neg}} = - \frac{1}{N_{\text{U}}} \sum_{i=1}^{N_{\text{U}}} \sum_{j=1}^{H \times W} \begin{cases}
    (1 - Y_{\text{L},ij}) \cdot \log \left( \frac{\exp(-P_{\text{$\theta$},ij})}{\sum_{k=1}^{H \times W} \exp(-P_{\text{$\theta$},ik})} \right), & \text{if } Y_{\text{L},ij} = 0 \text{ and } P_{\text{$\theta$},ij} < -t_{\text{neg}} \\
    0, & \text{otherwise}.
\end{cases}
\tag{3}
\end{equation}
Here, $N_{\text{U}}$ represents the total number of unlabeled images,
$P_{\text{$\theta$},ij}$  represent the logits predicted by the model, $\text{targets}$ are the ground truth labels, $Y_{\text{L},ij}$ is the ground truth label for the $i$-th sample and $j$-th class, and $H \times W$ represents the spatial dimensions of the image. 
The final Semi Cross Entropy Loss $\mathcal{L}_{\text{SemiCE}}$ is the sum of $\mathcal{L}_{\text{pos}}$ and $L_{\text{neg}}$:
The SemiCE loss is the same as the Consistency loss ($\mathcal{L}_{\text{cons}})$ for the student-teacher framework where this loss component tries to make the predictions of the student model consistent,
\begin{equation}
\mathcal{L}_{\text{cons}} = \mathcal{L}_{\text{SemiCE}}= \mathcal{L}_{\text{pos}} + \mathcal{L}_{\text{neg}}.
\tag{4}
\end{equation}
The overall loss function for the mean teacher network is a combination of the supervised loss and the consistency loss, weighted by respective coefficients. It can be represented in Equation 5 as:
\begin{equation}
\mathcal{L}_{\text{total}} = \mathcal{L}_{\text{sup}} + \lambda \mathcal{L}_{\text{cons}},
\tag{5}
\end{equation}
where $\lambda$ is the weighting coefficient that determines the importance of the consistency loss relative to the supervised loss. 
Section ~\ref{para:imbalance} further explains on the class imbalance selection of positive ($t_{\text{pos}}$)  and negative ($t_{\text{neg}}$) thresholds along with how different values of these threshold affect the results  

\subsection{Data Pre-Processing and Augmentations}

Data Pre-Processing and augmentations are crucial in enhancing our model’s ability to detect defects by introducing controlled variations into the training data. These variations enable the model to adapt to diverse real-world scenarios, capturing the nuances of defects under different conditions, thus contributing to improved accuracy and robustness in defect detection. Image pre-processing involves resizing and scaling pixel values as per the dataset and model requirements. They include normalizing images and converting images to tensors suitable for neural network input.  In our approach, we utilize both weak and strong augmentations to enrich the diversity of the training data.Weak augmentations are applied to the teacher model to generate better pseudo-labels. These subtle changes ensure the teacher model produces reliable pseudo-labels while being exposed to slight variations in the data. Strong augmentations are used for training the student model, exposing it to more challenging samples to enhance its ability to generalize from complex variations. They involve significant alterations to training images such as applying color jittering, random grayscale conversion and blurring the images, and also cut mix augmentation \cite{yun2019cutmix}. These augmentation techniques improve the overall performance and robustness of the model. Our semi-supervised learning framework and augmentation can be adapted to other datasets with the same pre-processing steps performed on the studied datasets. In this study, we have adapted the proposed model to two different datasets. This adaptability enhances the reproducibility and generalizability of our approach across various datasets.

\section{Experiments and Evaluation}
\label{sec:Results}











\paragraph{Datasets}
We use the UCF-EL Defect Dataset, which includes nine distinct defect types, each with specific EL patterns and locations on the cell surface, such as cracks, grid interruptions, and corrosion. The dataset also consists of images which are healthy and contain no defects. The ground truth segmentation map for these images has a pixel index same as the background. 

\begin{figure}[ht!]
\centering
\resizebox{0.9\textwidth}{!}{\includegraphics{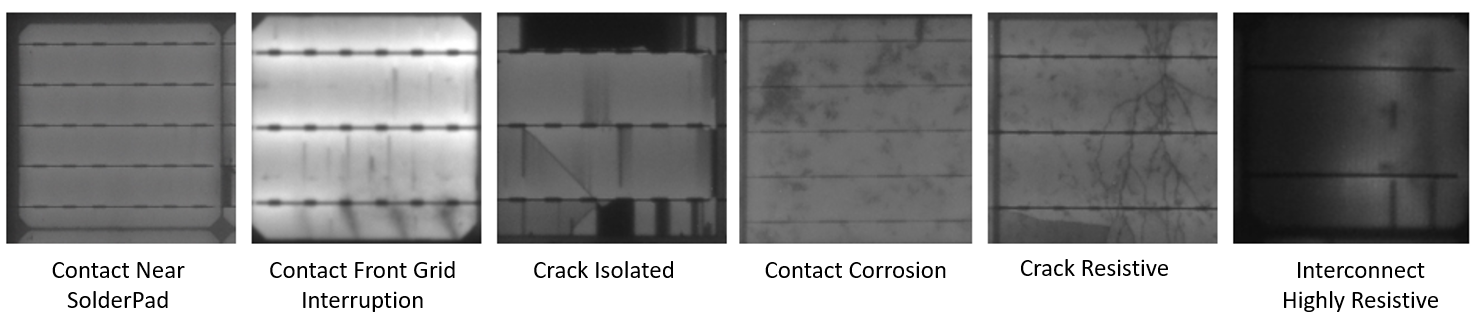}}
\caption{\textbf{Sample defect images from UCF-EL dataset:} 
The images from the left show defect types \enquote{Contact Near SolderPad}, \enquote{Contact Front Grid Interruption}, \enquote{Crack Isolated}, \enquote{Contact Corrosion}, \enquote{Crack Resistive} and \enquote{Interconnect Highly Resistive}. 
}
\label{fig:Defects}
\end{figure}

\begin{figure}[ht!]
\centering
\includegraphics[width=1.0\textwidth]{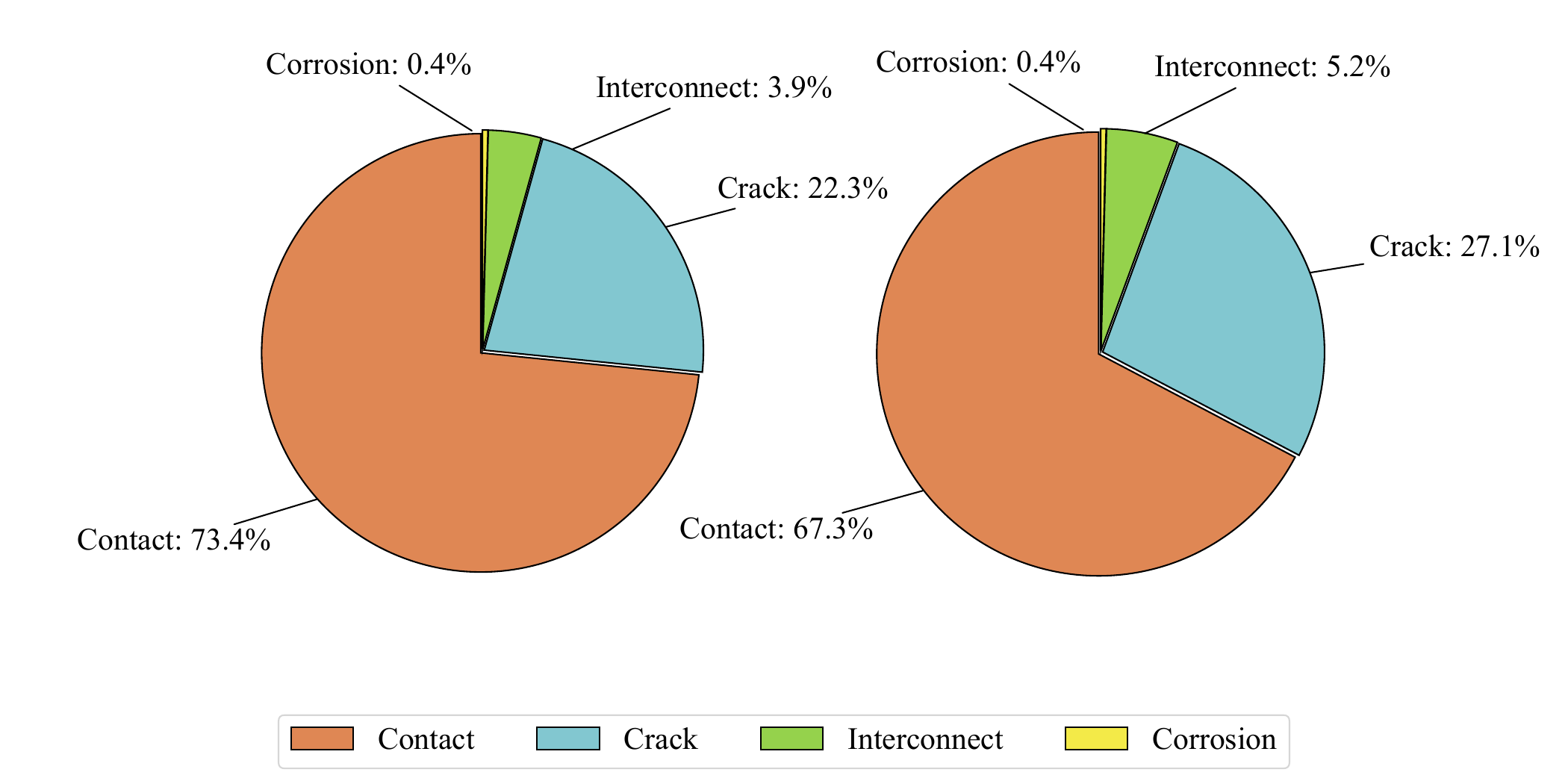}
\caption{\textbf{Distribution of defects for the UCF-EL dataset:} The pie chart on the left illustrates the distribution of the four defect classes based on the \emph{number of images} containing each defect class, indicating how frequently each type of defect appears in the dataset at the image level. In contrast, the pie chart on the right shows the \emph{pixel-wise} distribution of defects, detailing the proportion of \emph{pixels} occupied by each defect class across all images. This comparison highlights the difference between the occurrence of defects in images and their actual spatial extent within the dataset. 
}
\label{fig:piechart}
\end{figure}

\begin{figure}[ht!]
\centering
\includegraphics[height = 0.4\textwidth,width=1.0\textwidth]{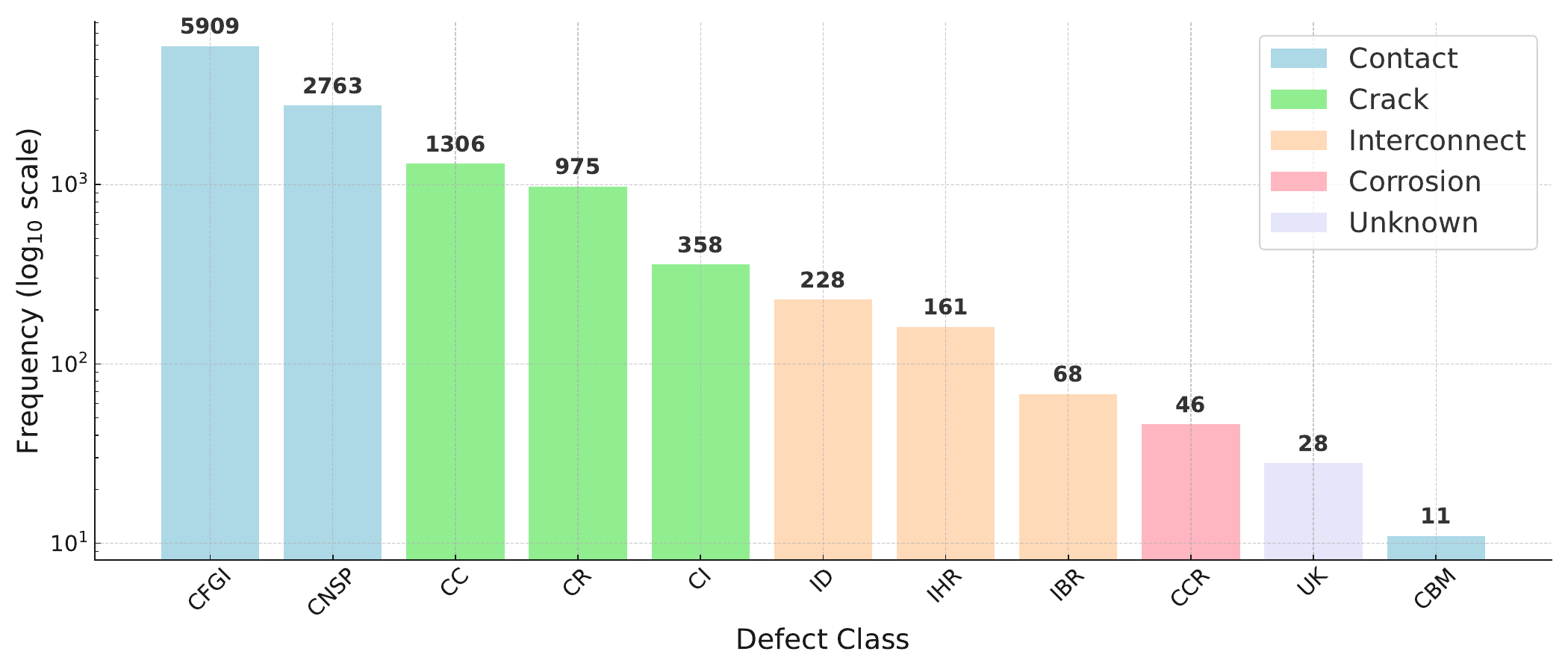}
\caption{
\textbf{Distribution of fine-grained defect classes for UCF-EL dataset:}
Each defect is denoted by an abbreviation for clarity. \enquote{CNSP} represents Contact Near Solder Pad, \enquote{CR} stands for Crack Resistive, \enquote{CFGI} is Contact Front Grid Interruption, \enquote{CC} denotes Crack Closed, \enquote{CI} indicates Crack Isolated, \enquote{IHR} means Interconnect Highly Resistive, \enquote{IBR} stands for Interconnect Bright Spot, \enquote{CCR} is Contact Corrosion, \enquote{UK} signifies Unknown defects, \enquote{CBM}represents Contact Belt Marks, and \enquote{ID} denotes Interconnect Disconnected.
}
\label{fig:DOD}
\end{figure}

In Figure ~\ref{fig:Defects}, we present a visual representation of some of the EL images containing defects. The defect are classified into four major defect classes which are Contact, Crack, Interconnect and Corrosion. Further each of these defects have sub categories for defect. 
Pie charts and histogram distributions of defect classes, as shown in Figures ~\ref{fig:piechart} and ~\ref{fig:DOD}, indicate Contact defects are the most prevalent, followed by Crack and Interconnect, while Corrosion defects are least common.

In addition to the primary UCF-EL dataset, our study uses different datasets to validate the model's performance evaluation and its ability to generalize across various scenarios. The study performed by \cite{pratt2023benchmark} used a combination of different datasets containing EL images from five different sources \cite{ karimi2019automated, Deitsch2019, Deitsch2021, Buerhop2018}. We call this combined dataset as Combined Solar Benchmark dataset (CSB-Dataset). This dataset is further divided into three subsets, Subset 1 (CSB-Dataset-S1) contains 593 images with 12 defect classes, Subset 2 (CSB-Dataset-S2) contains 2109 images with 15 defect classes and Subset 3 (CSB-Dataset-S3) contains 2282 images with 16 defect classes. Out of all these defect classes, the major defect classes are Crack, Gridline, and Inactive, and we focus on these three defect types in our experiments which also enables a fair comparison with \cite{pratt2023benchmark}.

\paragraph{Training Details}
\label{para:Training}
During the training phase, the model is trained with 20\% of the labeled data and 80\% of the unlabeled data from the training set. 
The training process is carried out for 30 epochs with batch size 16, Stochastic Gradient Descent (SGD) optimizer with learning rate 0.001, weight decay 0.0001 and momentum 0.9 on a 32GB GPU. Further consistency weight is taken as 1.5 and base size of image is \(299 \times 299\) and crop size of \(224 \times 224\). Further strong and weak augmentations are applied to student and teacher model. Strong augmentations include normalizing imags, color jittering (adjusting brightness, contrast, saturation, and hue by up to \(\pm50\%\)), random grayscale conversion with a 20\% probability, Gaussian blur with a random kernel size between 1 and 5 and sigma between 0.1 and 2.0, and \textit{CutMix augmentation} where patches of images are randomly replaced with patches from other images based on a beta distribution (\(\alpha=4\), \(\beta=4\)). Weak augmentations applied to teacher network only include normalizing the images.  The standard mean and standard deviation values of \([0.485, 0.456, 0.406]\) and \([0.229, 0.224, 0.225]\) respectively are used for normalizing the images for both student and teacher models. 


\paragraph{Evaluation Metrics}
\label{para:Metric}

To evaluate the performance of our proposed method, we use four different metrics: Mean Intersection over Union (mIoU), Precision, Recall, and F1 score. 
\textit{Precision} quantifies the proportion of predicted defect pixels that are correctly identified, and is defined as:
\begin{equation}
\text{Precision} = \frac{\text{TP}}{\text{TP} + \text{FP}},
\tag{6}
\end{equation}
where TP (true positives) denotes the number of correctly predicted defect pixels, and FP (false positives) denotes the number of non-defect pixels incorrectly predicted as defects.
\textit{Recall} measures the ability of the model to identify all actual defect pixels, and is given by:
\begin{equation}
\text{Recall} = \frac{\text{TP}}{\text{TP} + \text{FN}},
\tag{7}
\end{equation}
where FN (false negatives) corresponds to defect pixels that were not detected by the model.
\textit{F1 score}, the harmonic mean of precision and recall, offers a balanced metric that considers both false positives and false negatives:
\begin{equation}
\text{F1 Score} = 2 \times \frac{\text{Precision} \times \text{Recall}}{\text{Precision} + \text{Recall}}.
\tag{8}
\end{equation}
\textit{Mean IoU} (mIoU) evaluates the overlap between predicted segmentation masks and ground truth annotations, averaged across all classes. It is defined as:
\begin{equation}
\text{mIoU} = \frac{1}{N_{\text{classes}}} \sum_{i=1}^{N_{\text{classes}}} \frac{\text{TP}_i}{\text{TP}_i + \text{FP}_i + \text{FN}_i},
\tag{9}
\end{equation}
where \(N_{\text{classes}}\) is the total number of classes, and \(\text{TP}_i\), \(\text{FP}_i\), and \(\text{FN}_i\) represent the true positives, false positives, and false negatives for class \(i\), respectively.
These metrics provide a comprehensive assessment of segmentation accuracy, particularly in the context of defect detection in EL imagery.
Together, these metrics provide a comprehensive assessment for accuracy, completeness, and reliability of the segmentation model in identifying and localizing defects in photovoltaic modules.

\subsection{Quantitative Evaluation}
\label{subsec:Q_Results}

 Our quantitative analysis focused on the performance of PV-S3 with 20\% labeled data in defect detection. By comparing the model’s predictions with fully-supervised learning and the baseline model for supervised learning with 20\% labeled data along with semi-supervised learning with 100\% labeled data, we assess the model’s performance in identifying and segmenting defects  corresponding ground truths.

\paragraph{Comparison under limited labels}
 Regarding the performance on UCF-EL dataset with the semi-supervised learning scenario with 20\% labeled data, we record average values for metrics as  mIoU of 67\%, precision of 72.50\%, recall of 88.90\%, and F1-score of 78.17\% whereas with 20\% labeled data in the supervised learning yielded mIoU of 8.76\%, precision of 25.79\%, recall of 26.27\% and F1-score of 14.73\%. Based on the results we can observe that semi-supervised learning with PV-S3 has yielded significant improvement in results as compared to fully supervised learning with same amount of labeled dataset. The scores with the baseline model of supervised learning with 20\% labeled data and semi-supervised learning with 20\% labeled data has been summarized in Table ~\ref{tab:UCFel-Solar-SSL}.   
To further evaluate the robustness of PV-S3, we perform three experimental runs with different subsets of labeled samples which are randomly selected from the training dataset. The mean and standard deviation of performance metrics for each defect class are shown in Table ~\ref{tab: runs}. We observe that the proposed SSL approach significantly outperform the supervised baseline in all metrics consistently, demonstrating its robustness.

\begin{table}[ht!]
    \centering
    \Large
    \resizebox{\textwidth}{!}{
    \begin{tabular}{l|cc|cc|cc|cc}
        \toprule
        & \multicolumn{2}{c|}{mIoU} & \multicolumn{2}{c|}{Precision} & \multicolumn{2}{c|}{Recall} & \multicolumn{2}{c}{F1-Score} \\
        Defect Class & SL (20\%) & SSL (20\%) & SL (20\%) & SSL (20\%) & SL (20\%) & SSL (20\%) & SL (20\%) & SSL (20\%) \\
        \midrule
        Crack & 25.29 & 89.00 & 83.13 & 98.00 & 26.66 & 89.75 & 40.37 & 93.69 \\
        Contact & 6.89 & 49.00 & 16.73 & 57.00 & 10.49 & 79.50 & 12.89 & 66.39 \\
        Interconnect & 0.85 & 40.00 &  1.26 & 43.00 & 2.58 & 88.81 & 1.69 & 57.94 \\
        Corrosion & 2.03 & 90.00 & 2.05 & 92.00 & 65.37 & 97.54 & 3.98 & 94.68 \\ \hline
        Average & 8.76 & 67.00 & 25.79 & 72.50 & 26.27 & 88.90 & 14.73 & 78.17 \\
        \bottomrule
    \end{tabular}
    }
    \caption{\textbf{Comparison of PV-S3 under limited labels with our supervised method on UCF-EL dataset:} The table shows the performance of different defect classes under Supervised Learning (SL) with 20\% labeled images and Semi-Supervised Learning (SSL) with 20\% labeled images. The metrics evaluated include IoU, Precision, Recall, and F1-Score, facilitating a direct comparison of the impact of labeling extent on model performance.}
    \label{tab:UCFel-Solar-SSL}
\end{table}

\begin{table}[ht!]
    \centering
    \Large
    \resizebox{\textwidth}{!}{%
    \begin{tabular}{l|cc|cc|cc|cc}
        \toprule
        & \multicolumn{2}{c|}{mIoU} & \multicolumn{2}{c|}{Precision} & \multicolumn{2}{c|}{Recall} & \multicolumn{2}{c}{F1-Score} \\
        Defect Class & SL (20\%) & SSL (20\%) & SL (20\%) & SSL (20\%) & SL (20\%) & SSL (20\%) & SL (20\%) & SSL (20\%) \\
        \midrule
        Crack & $26.20 \pm 0.66$ & $89.05 \pm 0.38$ & $83.50 \pm 1.01$ & $97.97 \pm 0.24$ & $26.66 \pm 0.64$ & $90.43 \pm 0.78$ & $41.41 \pm 0.73$ & $94.05 \pm 0.33$ \\
        Contact & $6.95 \pm 0.06$ & $49.95 \pm 1.02$ & $16.92 \pm 0.78$ & $58.52 \pm 1.71$ & $11.06 \pm 1.55$ & $77.16 \pm 1.17$ & $12.67 \pm 0.89$ & $66.95 \pm 0.92$ \\
        Interconnect & $0.7 \pm 0.26$ & $41.05 \pm 0.85$ & $1.03 \pm 0.38$ & $42.86 \pm 0.81$ & $2.13 \pm 0.78$ & $92.07 \pm 2.90$ & $1.39 \pm 0.51$ & $58.46 \pm 0.55$ \\
        Corrosion & $2.48 \pm 0.61$ & $86.09 \pm 7.37$ &  $2.02 \pm 0.06$ & $92.62 \pm 2.02$ & $62.40 \pm 4.86$ & $92.24 \pm 6.99$ & $3.92 \pm 0.13$ & $92.33 \pm 4.36$ \\ \hline
        Average & $9.06 \pm 0.09$ & $66.53 \pm 1.35$ & $25.86 \pm 0.05$ & $72.99 \pm 0.35$ & $26.53 \pm 1.43$ & $88.17 \pm 1.94$ & $14.85 \pm 0.05$ & $78.00 \pm 0.78$ \\
        \bottomrule
    \end{tabular}%
    }
    \caption{\textbf{Robustness of PV-S3 with variation on data sampling:} Three runs are conducted for supervised learning with 20\% labeled images and semi-supervised learning with 20\% labeled images, and mean and standard deviation are recorded for performance metrics for each defect class along with their average values. This assessment allows for direct comparison between SL and SSL approaches for each metric.}
    \label{tab: runs}
\end{table}

\paragraph{Comparison with fully labeled dataset}

With 100\% labeled data using fully supervised learning, the Deeplabv3+ model achieves mean metrics of mIoU: 70.84\%, precision: 80.12\%, recall: 86.29\%, and F1-score: 82.03\% (Table ~\ref{tab:UCFel-Solar-FSL}). Comparing these results with our 20\% labeled semi-supervised learning approach, PV-S3 yields comparable results to fully supervised learning, with even higher precision. Leveraging 100\% labeled data in the semi-supervised framework with PV-S3 significantly improves performance, yielding mean mIoU: 74.75\%, precision: 79.96\%, recall: 91.30\%, and F1-score: 84.24\%. This improvement underscores the benefits of a larger dataset in better understanding and segmenting defects in PV module EL images.

\begin{table}[ht!]
    \centering
    \Large
    \resizebox{\textwidth}{!}{
    \begin{tabular}{l|cc|cc|cc|cc}
        \toprule
        & \multicolumn{2}{c|}{mIoU} & \multicolumn{2}{c|}{Precision} & \multicolumn{2}{c|}{Recall} & \multicolumn{2}{c}{F1-Score} \\
        Defect Class & FSL (100\%) & SSL (100\%) & FSL (100\%) & SSL (100\%) & FSL (100\%) & SSL (100\%) & FSL (100\%) & SSL (100\%) \\
        \midrule
        Crack & 88.38 & 91.78 &  96.91 & 97.54& 90.94 & 93.96 & 93.83 & 95.72 \\
        Contact & 52.92 & 58.29 & 60.56 & 69.20 & 80.75 & 78.70 & 69.21 & 73.65 \\
        Interconnect & 59.70 & 53.89 & 63.59 & 55.49& 90.70 & 94.93 & 74.76 & 70.04 \\
        Corrosion & 82.39 & 95.26 & 99.43 & 97.61 & 82.78 & 97.62 & 90.34 & 97.57 \\ \hline
        Average & 70.84 & 74.75 & 80.12 & 79.96 & 86.29 & 91.30 & 82.03 & 84.24 \\
        \bottomrule
    \end{tabular}
    }
    \caption{\textbf{Comparison of PV-S3 with our fully supervised method on UCF-EL dataset:} The table shows the performance of different defect classes under Semi Supervised Learning (SSL) with 100\% labeled images, and Fully Supervised Learning (FSL) with 100\% labeled images. The metrics evaluated include mIoU, Precision, Recall, and F1-Score, facilitating a direct comparison of the impact of labeling extent on model performance.}
    \label{tab:UCFel-Solar-FSL}
\end{table}

\begin{table}
  \centering 
  \small
  \begin{tabular}{@{}l | S[table-format=3.0] | S[table-format=2.2] S[table-format=2.2] S[table-format=2.2] S[table-format=2.2]@{}}
    \toprule
    Baselines & {Labeled Data (\%)} & {mIoU (\%)} & {Precision (\%)} & {Recall (\%)} & {F1-Score (\%)} \\
    \midrule
    Deeplabv3 \cite{fioresi2021automated} & 100 & 57.30 &59.00 & 59.75 & 57.75 \\
    {Deeplabv3+ (Ours FSL)} & 100 & 70.84 & \textbf{80.12} & 86.29 & 82.03 \\ 
    {PV-S3 (Ours-SSL (20\%))} & 20 & 67.00 & 72.50 & 88.90 & 78.17 \\
    {PV-S3 (Ours-SSL (100\%))} & 100 & \textbf{74.75} & 79.96 & \textbf{91.30} & \textbf{84.24} \\
    \bottomrule
  \end{tabular}
  \caption{\textbf{Comparison of PV-S3 with existing supervised approaches on UCF-EL dataset:} 
  Fully Supervised Learning with 100\% labeled data using Deeplabv3/Deeplabv3+ and Semi-Supervised Learning using Deeplabv3+ (Ours-SSL(20\%)) with 20\% labeled images and Semi-Supervised Learning using Deeplabv3+ with 100\% labeled data (Ours-SSL(100\%)). PV-S3 performed  better in terms of all four key metrics significantly reducing the annotation cost.
  }
  \label{tab:ucf-el}
\end{table}

Comparisons with existing Fully Supervised learning on UCF-EL dataset, performed by \cite{fioresi2021automated}, across key metrics are summarized in Table ~\ref{tab:ucf-el}. We observe improved performance with merely 20\% labeled samples and significant enhancement in mIoU, Precision, and F1-Score using all labeled samples. Our approach, utilizing just 20\% of labeled samples, yields a notable absolute enhancement, including a 9.7\% increase in mIoU, a 13.5\% rise in Precision, a 29.15\% boost in Recall, and a 20.42\% improvement in F1-Score compared to the previous state-of-the-art supervised method, reducing annotation expenses by 80\%.

\begin{table}[ht!]
    \centering    
    \Large
    \resizebox{\columnwidth}{!}{
    \begin{tabular}{c|c|cccccccc} 
         \toprule
         \multicolumn{2}{c|}{Experiment} & \multicolumn{2}{c}{mIoU (\%)} & \multicolumn{2}{c}{Precision (\%)} & \multicolumn{2}{c}{Recall (\%)} & \multicolumn{2}{c}{F1-Score (\%)} \\
         \cmidrule(lr){1-2} \cmidrule(lr){3-4} \cmidrule(lr){5-6} \cmidrule(lr){7-8} \cmidrule(lr){9-10}
         Dataset & Defect Type & SSL (20\%) & FSL (100\%) & SSL (20\%) & FSL (100\%) & SSL (20\%) & FSL (100\%) & SSL (20\%) & FSL (100\%) \\
         \midrule
         \multirow{4}{*}{CSB-Dataset-S1} & Crack & $50.00$ & $58.00$  & $80.00$ & $86.00$ & $57.00$ & $64.00$  & $66.00$ & $73.00$ \\
         & Gridline & $74.00$ & $83.00$  & $88.00$ & $91.00$ & $82.00$ & $91.00$  & $85.00$ & $91.00$ \\
         & Inactive & $50.00$ & $50.00$  & $66.00$ & $62.00$ & $68.00$ & $71.00$  & $67.00$ & $66.00$ \\ \cmidrule(lr){2-10}
         & Average & $58.00$ & $63.67$ & $78.00$ & $79.67$ & $69.00$ & $75.33$ & $72.67$ & $76.67$ \\
         \cmidrule(lr){1-10}
         \multirow{4}{*}{CSB-Dataset-S2} & Crack & $61.00$ & $56.00$ & $74.00$ & $72.00$ & $78.00$ & $71.00$ & $76.00$ & $71.00$ \\
         & Gridline & $78.00$ & $84.00$ & $88.00$ & $91.00$ & $87.00$ & $92.00$ & $87.00$ & $91.00$ \\
         & Inactive & $62.00$ & $60.00$ & $86.00$ & $77.00$ & $70.00$ & $73.00$ & $77.00$ & $75.00$ \\ \cmidrule(lr){2-10}
         & Average & $67.00$ & $66.67$ & $82.67$ & $80.00$ & $78.33$ & $78.67$ & $80.00$ & $79.00$ \\ 
         \cmidrule(lr){1-10}
         \multirow{4}{*}{CSB-Dataset-S3} & Crack & $55.00$ & $60.00$ & $62.00$ & $76.00$ & $84.00$ & $73.00$ & $71.00$ & $75.00$ \\
         & Gridline & $77.00$ & $85.00$ & $86.00$ & $93.00$ & $88.00$ & $90.00$ & $87.00$ & $92.00$ \\
         & Inactive & $62.00$ & $60.00$ & $86.00$ & $90.00$ & $65.00$ & $87.00$ & $75.00$ & $83.00$ \\ \cmidrule(lr){2-10}
         & Average & $64.67$ & $68.33$ & $78.00$ & $86.33$ & $79.00$ & $83.33$ & $77.67$ & $83.33$ \\ 
         \bottomrule
    \end{tabular}
    } 
    \caption{\textbf{Comparison with our supervised baselines on CSB-Dataset:} Here we use the PV-S3 on Combined Solar Benchmark dataset (CSB-Dataset) to compare the scores of semi-supervised learning with fully supervised learning (FSL) across four key metrics.
    }
    \label{tab: Solar_23 Comparisons}
\end{table}

\paragraph{Comparison on Combined Solar Benchmark Dataset}
On the \textbf{C}ombined \textbf{S}olar \textbf{B}enchmark Dataset (CSB-Dataset) we evaluate PV-S3 on all three subsets with three major defect classes Crack, Gridline and Inactive. The results for four key metrics, mIoU, Precision, Recall and F1-Score, are summarized in Table ~\ref{tab: Solar_23 Comparisons}. We observe competitive performance with PV-S3 using only 20\% labeled samples across all metrics when compared with fully supervised method using 100\% labeled samples. 
We also compare the performance of PV-S3 with existing methods on these datasets. This comparison is shown in Table  ~\ref{tab:baselines}.
The performance of PV-S3 is compared with various baseline models which include PSPNet, UNet-12 and UNet-25 using both 20\% (Our-SSL) and 100\% (Our-FSL) labels. 
We observe that conventional semantic segmentation models like PSPNet and UNet variants which are UNet-12 and UNet-25 performed poorly in terms of the Intersection of Union as compared to Deeplab variants. In terms of recall score, UNet-25 has yielded the highest recall score of 60.66\% while PSPNet yielded the lowest score of 43.0\%.  We observe that PV-S3 outperforms other baseline models in terms of all metric scores. Using merely 20\% of labeled images for training we observed an increase of 30\% in IoU with a reduction in 80\% annotation cost compared to the state-of-the-art model on CSB-Dataset, further using our model with 100\% labeled images an increase of 35.67\% in mIoU and 0.33\% increase in recall was observed with state-of-the-art model on CSB-Dataset.





\begin{table}[ht!]
  \centering 
  \small
  \begin{tabular}{@{}l S[table-format=3.0] S[table-format=2.2] S[table-format=2.2] S[table-format=2.2] S[table-format=2.2]@{}}
    \toprule
    Baseline Models & {Labeled Data (\%)} & {mIoU (\%)} & {Precision (\%)} & {Recall (\%)} & {F1-Score (\%)} \\
    \midrule
    PSPNet \cite{pratt2023benchmark} & 100 & 14.00 & \multicolumn{1}{c}{-} & 43.00 & \multicolumn{1}{c}{-} \\
    UNet-12 \cite{pratt2021defect} & 100 & 16.00 & \multicolumn{1}{c}{-} & 58.00 & \multicolumn{1}{c}{-} \\
    UNet-25 \cite{pratt2021defect} & 100 & 13.33 & \multicolumn{1}{c}{-} & 60.67 & \multicolumn{1}{c}{-} \\
    Deeplabv3+ \cite{pratt2023benchmark} & 100 & 28.00 & \multicolumn{1}{c}{-} & 75.00 & \multicolumn{1}{c}{-} \\
    \rowcolor{gray!25}{PV-S3} (Ours-FSL) & 100 & \textbf{63.67} & \textbf{79.67} & \textbf{75.33} & \textbf{76.67} \\ 
    \hline
    \rowcolor{gray!25}{PV-S3 (Ours-SSL)} & 20 & 58.00 & 78.00 & 69.00 & 72.67 \\
    \bottomrule
  \end{tabular}
  \caption{\textbf{Comparison with existing supervised methods on CSB-Dataset-S1:} We compare the results of PV-S3 with different variations with other baselines for the CSB-Dataset-S1. Deeplabv3+ with 100\% labeled data (Ours-FSL) and Deeplabv3+ (Ours-SSL) with 20\% labeled images performed significantly better in terms of IoU and Recall scores.
  }
  \label{tab:baselines}
\end{table}

\subsection{Discussion and Analysis}

We further study some key aspects
such as ablation study of loss function, effectiveness of Semi Cross-Entropy (SemiCE) loss in addressing class imbalance, impact of variations in the amount of labeled data on our model's performance, and the confidence level of our model in segmenting defects. 

\paragraph{Ablation Study for Loss Function}

We conducted an ablation study on the UCF-EL dataset to compare the performance of the proposed Semi Cross-Entropy (SemiCE) loss with the standard Mean Squared Error (MSE) loss, as shown in Table~\ref{tab:ablation}. While the overall results with MSE loss are comparable to those obtained using the SemiCE loss, we observed significant differences in the detection of certain defect classes, especially those with class imbalances like \textit{Interconnect} and \textit{Contact} defects.
The MSE loss treats all defect classes uniformly during loss calculation, which can be suboptimal for underrepresented classes. In contrast, the SemiCE loss incorporates positive and negative thresholds, allowing the model to optimize predictions for each defect class differently. This approach enhances learning by giving appropriate emphasis to classes with fewer samples. Our results indicate that using MSE loss led to lower performance on \textit{Interconnect} and \textit{Contact} defects due to its inability to effectively handle class imbalance. However, when using the SemiCE-S2 variant of our loss function, we observed improved performance on these defect classes. This demonstrates the effectiveness of the SemiCE loss in addressing class imbalance and enhancing overall detection accuracy.
\begin{table}[ht!]
    \centering
    \Large
    \resizebox{\textwidth}{!}{
    \begin{tabular}{l|ccc|ccc|ccc|ccc}
        \toprule
        & \multicolumn{3}{c|}{mIoU} & \multicolumn{3}{c|}{Precision} & \multicolumn{3}{c|}{Recall} & \multicolumn{3}{c}{F1-Score} \\
        Defect Class & MSE & SemiCE-S1 & SemiCE-S2 & MSE & SemiCE-S1 & SemiCE-S2 & MSE & SemiCE-S1 & SemiCE-S2 & MSE & SemiCE-S1 & SemiCE-S2 \\
        \midrule
        Crack & \textbf{91.00} & 89.00 & 89.00 &  97.35 & \textbf{98.00} & 96.00 & 91.27 & 89.75 & \textbf{92.00} & \textbf{94.21} & 93.69 & 94.00 \\
        Contact & 51.00 & 49.00 & \textbf{53.00} & 59.51 & 57.00 & \textbf{63.00} & 77.55 & \textbf{79.50} & 78.00 & 68.23 & 66.39 & \textbf{70.00} \\
        Interconnect & 45.00 & 40.00 & \textbf{64.00} & 49.56 & 43.00 & \textbf{74.00} & 82.71 & \textbf{88.81} & 83.00 & 63.54 & 57.94 & \textbf{78.00} \\
        Corrosion & 86.00 & \textbf{90.00} & 81.00 & \textbf{96.95} & 92.00 & 91.00 & 87.96 & \textbf{94.68} & 88.00 & 92.23 & \textbf{97.57} & 89.00 \\ \hline
        Average & 67.75 & 67.00 & \textbf{71.75} & 75.84 & 72.50 & \textbf{81.00} & 84.87 & \textbf{88.90} & 85.25 & 79.65 & 78.17 & \textbf{82.75}\\
        \bottomrule
    \end{tabular}
    }
    \caption{\textbf{Ablation study of SemiCE loss function with MSE loss function for PV-S3 model:} The table shows ablation study conducted for SemiCE loss function for different threshold with Mean Squared Error (MSE) loss function. Two setting representing different positive and negative threshold is taken, SemiCE-S1 represents the first setting with positive threshold of 0.6 and negative threshold of 0 and SemiCE-S2 represents the second setting with positive threshold of 0.2 and negative threshold of 0.4.}
    \label{tab:ablation}
\end{table}
\paragraph{Analysing SemiCE Loss for Class Imbalance}
\label{para:imbalance}

Detecting defects in PV modules presents a significant challenge due to class imbalance: certain defects, such as \textit{Interconnect} and \textit{Corrosion}, occur less frequently in the dataset compared to defects like \textit{Crack} and \textit{Contact}. To address this imbalance, we employ the Semi Cross-Entropy (SemiCE) loss function, which uses adaptive thresholds for positive and negative predictions during training. The positive threshold ($t_{pos}$) controls how confidently a prediction must be classified as a defect, while the negative threshold ($t_{neg}$) ensures that only strong negative predictions contribute to the loss. This threshold-based filtering helps counter the common issue in imbalanced datasets where the model tends to overpredict frequent classes and underpredict rare ones due to skewed confidence distributions. By lowering $t_{pos}$, the model becomes more receptive to low-confidence predictions, improving recall for minority classes. Conversely, a higher $t_{pos}$ enforces stricter confidence, favoring precision. The negative threshold further ensures that noisy or ambiguous negatives do not dominate the learning signal, helping stabilize training across all classes, including those with fewer examples.

In our experiments, we varied both the SemiCE positive and negative thresholds to observe their impact on Interconnect defect class performance. Performance with different thresholds is shown in Table~\ref{tab:threshold}, with original results in Table ~\ref{tab:UCFel-Solar-SSL} using initial thresholds of 0.6 (positive) and 0 (negative), chosen for their balanced nature.
We conducted experiments by varying the thresholds to observe their impact on model performance across multiple defect types, changing the thresholds significantly affects not only Interconnect defects but also Crack, Contact, and Corrosion classes. For instance, a lower positive threshold ($t_{pos} = 0.2$) and moderate negative threshold ($t_{neg} = 0.4$) yielded improved performance across all defect classes, with Interconnect achieving an mIoU of 64\% and notable improvements in F1-scores for Crack and Contact defects as well.
Through experimentation, we aimed to identify threshold values optimizing the metrics while maintaining model consistency, shedding light on the complex relationship between prediction confidence and model performance.

These observations suggest that adjusting thresholds helps the model better capture difficult defects (such as Interconnect) by improving performance, while still maintaining high performance for more frequent classes (such as Crack). Thus, the SemiCE loss function, through careful threshold tuning, effectively mitigates problem of detecting difficult defects across different defect types in the dataset. Based on our experiments, we chose the thresholds ($t_{pos} = 0.2$, $t_{neg} = 0.4$) to achieve the best overall balance between precision and recall, while maintaining model consistency.

\begin{table}[ht!]
    \centering
    \setlength{\tabcolsep}{4pt} 
    \renewcommand{\arraystretch}{1.1} 
    \scriptsize 
    \resizebox{\textwidth}{!}{ 
    \begin{tabular}{c|cccc|cccc|cccc|cccc|cccc}
        \hline
        Threshold & \multicolumn{4}{c|}{Crack} & \multicolumn{4}{c|}{Contact} & \multicolumn{4}{c|}{Interconnect} & \multicolumn{4}{c|}{Corrosion} & \multicolumn{4}{c}{Average} \\
        \cline{2-21}
         Pos, Neg & mIoU & Prec & Rec & F1 & mIoU & Prec & Rec & F1 & mIoU & Prec & Rec & F1 & mIoU & Prec & Rec & F1 & mIoU & Prec & Rec & F1 \\
        \hline
        0.0, 0.0 & 89 & 95 & 93 & 94 & 51 & 65 & 71 & 68 & 50 & 71 & 65 & 68 & 92 & 98 & 94 & 96 & 71 & 82 & 81 & 82 \\
        \rowcolor{green!25} 0.2, 0.4  & 89 & 96 & 92 & 94 & 53 & 63 & 78 & 70 & 64 & 74 & 83 & 78 & 81 & 91 & 88 & 89 & 72 & 81 & 85 & 83 \\
        0.3, 0.2 & 89 & 98 & 91 & 94 & 52 & 61 & 77 & 68 & 44 & 50 & 79 & 61 & 67 & 69 & 96 & 80 & 63 & 70 & 86 & 76 \\
        0.4, 0.2 & 91 & 97 & 93 & 95 & 52 & 64 & 73 & 68 & 45 & 48 & 81 & 62 & 76 & 77 & 98 & 86 & 66 & 72 & 86 & 78 \\
        0.5, 0.5 & 91 & 98 & 93 & 95 & 53 & 64 & 75 & 69 & 46 & 46 & 97 & 63 & 85 & 87 & 97 & 92 & 69 & 74 & 91 & 80 \\
        0.6, 0.0 & 89 & 98 & 90 & 94 & 49 & 57 & 79 & 66 & 40 & 43 & 89 & 58 & 90 & 92 & 97 & 95 & 67 & 73 & 89 & 78 \\
        \hline
    \end{tabular}
    } 
    \caption{\textbf{Variation of performance across different thresholds:} This table shows the percentages for IoU, Precision, Recall, and F1-Score for the defect classes of Crack, Contact, Interconnect, and Corrosion across different thresholds. Best overall performance of the model is highlighted with green.}
    \label{tab:threshold}
\end{table}

\begin{figure}[ht!]
\centering
\includegraphics[width=0.15\textwidth, height=0.15\textwidth]{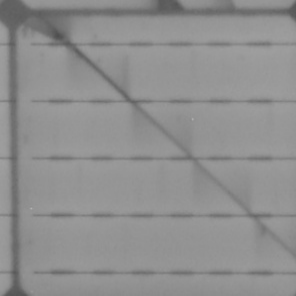}
\includegraphics[width=0.15\textwidth, height=0.15\textwidth]{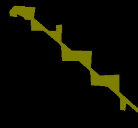}
\includegraphics[width=0.15\textwidth, height=0.15\textwidth]{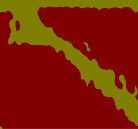}
\quad
\includegraphics[width=0.15\textwidth, height=0.15\textwidth]{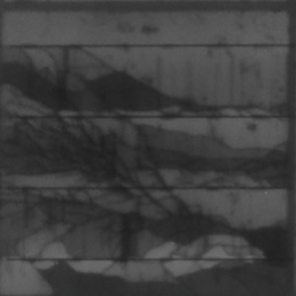}
\includegraphics[width=0.15\textwidth, height=0.15\textwidth]{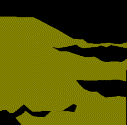}
\includegraphics[width=0.15\textwidth, height=0.15\textwidth]{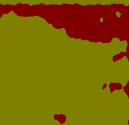}
\\ \vspace{5pt}
\includegraphics[width=0.15\textwidth, height=0.15\textwidth]{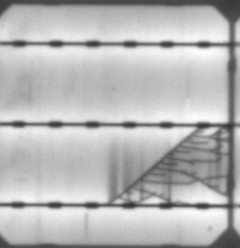}
\includegraphics[width=0.15\textwidth, height=0.15\textwidth]{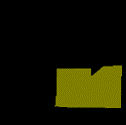}
\includegraphics[width=0.15\textwidth, height=0.15\textwidth]{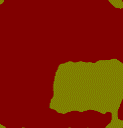}
\quad
\includegraphics[width=0.15\textwidth, height=0.15\textwidth]{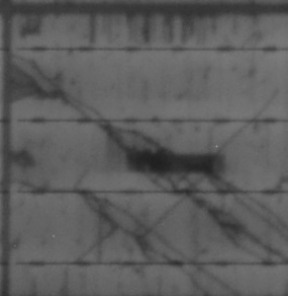}
\includegraphics[width=0.15\textwidth, height=0.15\textwidth]{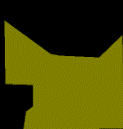}
\includegraphics[width=0.15\textwidth, height=0.15\textwidth]{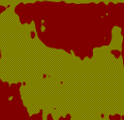}
\caption{
\textbf{Visualizing semantic segmentation:} 
Here we show visual analysis of the results obtained by PV-S3 for various defect categories. In each set of three images, the first image shows EL image, the second image is the ground-truth segmentation for corresponding defect and the third image shows the segmentation provided by PV-S3.
}
\label{fig: GT Vs Original}
\end{figure}

\begin{figure}[ht!]
\centering
\includegraphics[width=0.15\textwidth, height=0.15\textwidth]{images/predictions/el1.png}
\includegraphics[width=0.15\textwidth, height=0.15\textwidth]{images/predictions/gt1.png}
\includegraphics[width=0.15\textwidth, height=0.15\textwidth]{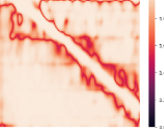}
\quad
\includegraphics[width=0.15\textwidth, height=0.15\textwidth]{images/predictions/el2.png}
\includegraphics[width=0.15\textwidth, height=0.15\textwidth]{images/predictions/gt2.png}
\includegraphics[width=0.15\textwidth, height=0.15\textwidth]{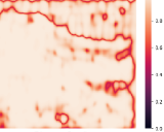}
\\ \vspace{5pt}
\includegraphics[width=0.15\textwidth, height=0.15\textwidth]{images/predictions/el3.png}
\includegraphics[width=0.15\textwidth, height=0.15\textwidth]{images/predictions/gt3.png}
\includegraphics[width=0.15\textwidth, height=0.15\textwidth]{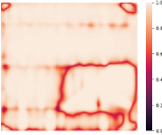}
\quad
\includegraphics[width=0.15\textwidth, height=0.15\textwidth]{images/predictions/el4.png}
\includegraphics[width=0.15\textwidth, height=0.15\textwidth]{images/predictions/gt4.png}
\includegraphics[width=0.15\textwidth, height=0.15\textwidth]{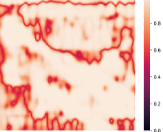}
\caption{
\textbf{Analyzing models confidence:}
Here we analyse the confidence of PV-S3 for defect segmentation and observe a lower confidence in boundary regions of detected defects. In each set of three images, the first image is the EL image, the second image shows ground-truth segmentation mask, and the third image shows confidence of models prediction.}
\label{fig: cm1}
\end{figure}

\paragraph{Analysing Confidence in Segmentation} 
The model accurately identifies defects within regions but struggles near boundaries, as shown in Figure ~\ref{fig: GT Vs Original}. This highlights challenges in precise boundary detection, necessitating improved boundary-aware models. While overall accuracy is good, refinement is needed for handling intricate defect details, enhancing reliability in solar PV defect detection. Our analysis of the confidence map reveals its importance in understanding prediction certainty. Overlaying it with ground truth labels and EL images enables comprehensive evaluation, pinpointing areas for improvement. Lower confidence scores around boundaries indicate detection challenges, while higher scores elsewhere demonstrate effectiveness in identifying defects. Figure ~\ref{fig: cm1} illustrates the confidence map's visual representation of prediction certainty, with higher scores indicating greater confidence.

\begin{figure}[t!]
\centering

\includegraphics[width=1.0\textwidth]{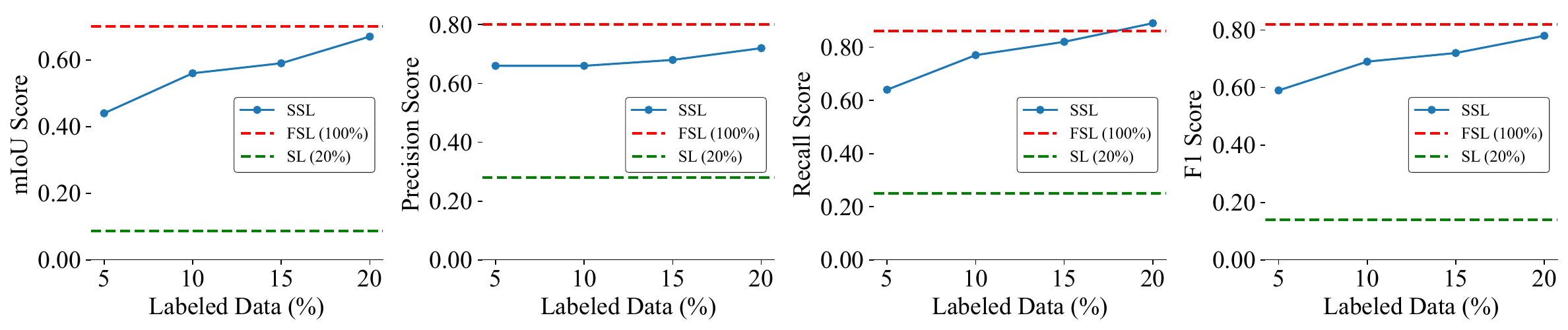}
\caption{
\textbf{Performance variation with amount of labels:}  
The plots show variation in models performance as we increase the amount of labeled samples from 5\% to 20\%. We observe that with merely 20\% of labeled samples, the model is able to provide performance comparable to fully supervised method which use 100\% labels. SSL- proposed PV-S3 approach for semi-supervised learning, SL- supervised learning with 20\% labels, and FSL- fully supervised method using 100\% labels.
}
\label{fig: Metrics Vs labeled data }
\end{figure}


\paragraph{Effect of Variation of Labeled Data in Training} 

We evaluated PV-S3's performance using different proportions of labeled data (5\% to 20\%) and compared it with fully supervised learning using 100\% labeled data. Results in Figure ~\ref{fig: Metrics Vs labeled data } indicate that semi-supervised learning with 5\% labeled data initially underperforms compared to fully supervised learning. However, performance notably improves at 10\% labeled data and continues to increase gradually up to 20\%, approaching scores achieved with 100\% labeled data for mIoU, Recall, and F1 score. Notably, recall scores for semi-supervised learning even exceed those of fully supervised learning. 
Further we observe that the recall score for the semi-supervised model trained with 20\% labeled data and 80\% unlabeled data exceeds that of the model trained with 100\% labeled data. Usually, semi-supervised learning methods which utilized unlabelled samples outperform fully supervised methods \cite{xie2020unsupervised, sohn2020fixmatch, xie2020self}, This counterintuitive result can be attributed to the benefits of semi-supervised learning in leveraging unlabeled data to improve model generalization. The inclusion of unlabeled data acts as a regularizer, preventing the model from overfitting to the labeled data, which might contain noise or biases. In contrast, training on 100\% labeled data without the additional unlabeled data may lead to overfitting, especially if the labeled data is limited in diversity.By incorporating a large amount of unlabeled data, the semi-supervised model learns more robust feature representations, capturing the underlying structure of the data more effectively. The semi-supervised approach allows the model to generalize better to unseen data, resulting in higher recall by correctly identifying more true positive instances of defects.


\paragraph{Analysing Model Training}

\begin{figure}[ht!]
\centering
\includegraphics[width=1.0\textwidth]{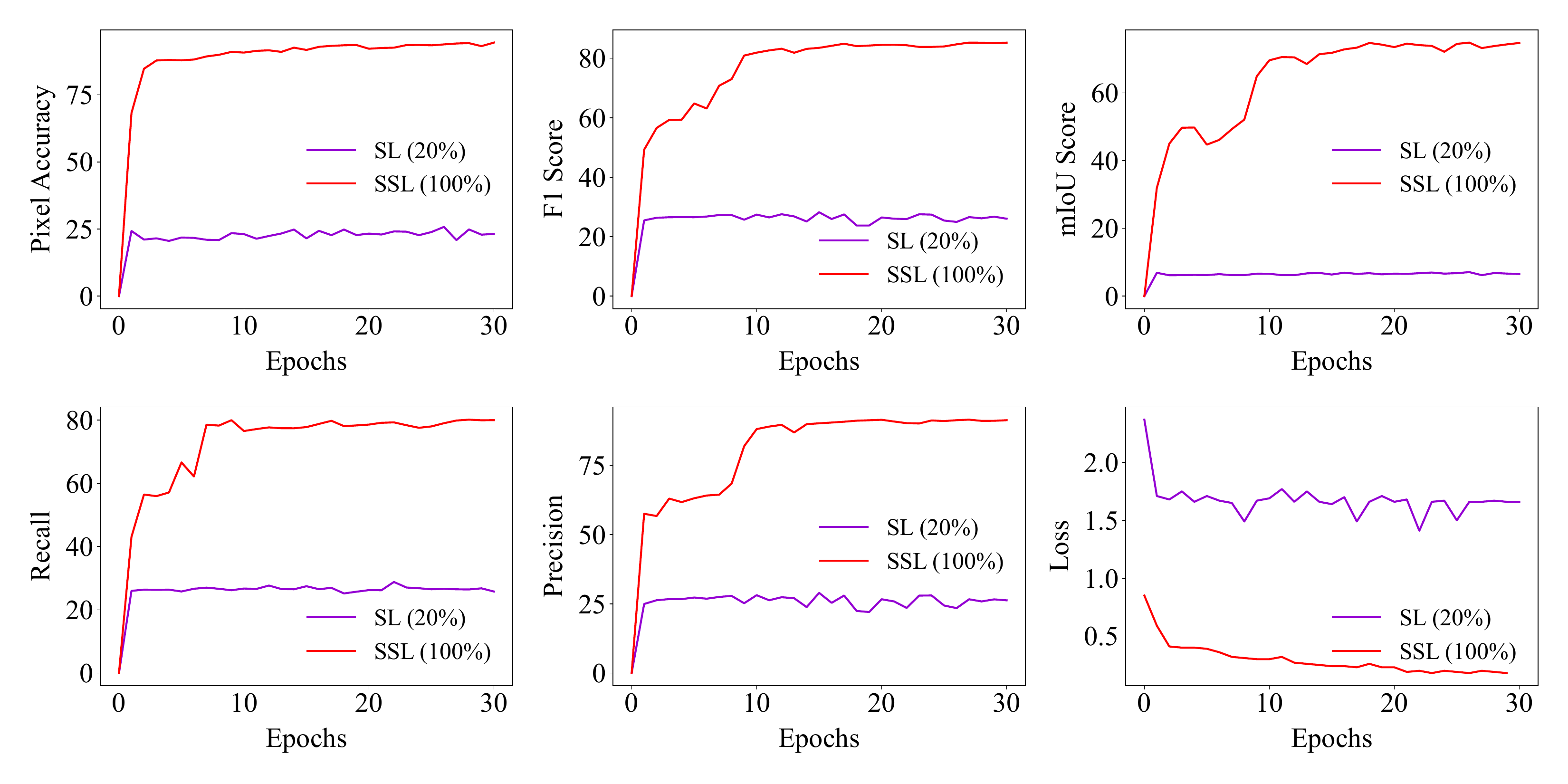}
\caption{
\textbf{Analyzing model training:}  
The top row plots show the variation of pixel accuracy (left), loss (middle), and mIoU (right) with varying training epochs, and the bottom row shows the performance metrics variation with epochs for supervised and semi-supervised learning with 20\% images for precision (left), recall (middle), and F1-score (right).}  

\label{fig: TRAINING CURVES}
\end{figure}

We analyze the training process of PV-S3 with epochs and compare with the supervised approach.
The plots for different metrics are shown in Figure ~\ref{fig: TRAINING CURVES}. Here we observe that for supervised learning with only 20\% images (SL (20\%)), due to less number of images and class imbalance, the model is not able to learn the features accurately and hence it has very low pixel accuracy and high loss values and have low values of precision and recall as well. 
For PV-S3, we observe an increasing trend for pixel accuracy, precision, and recall and decreasing trend for loss as training progresses.

\paragraph{Analyzing Confusion Matrix}

\begin{figure}[ht!]
\centering
  \centering
  \includegraphics[width=0.75\textwidth]{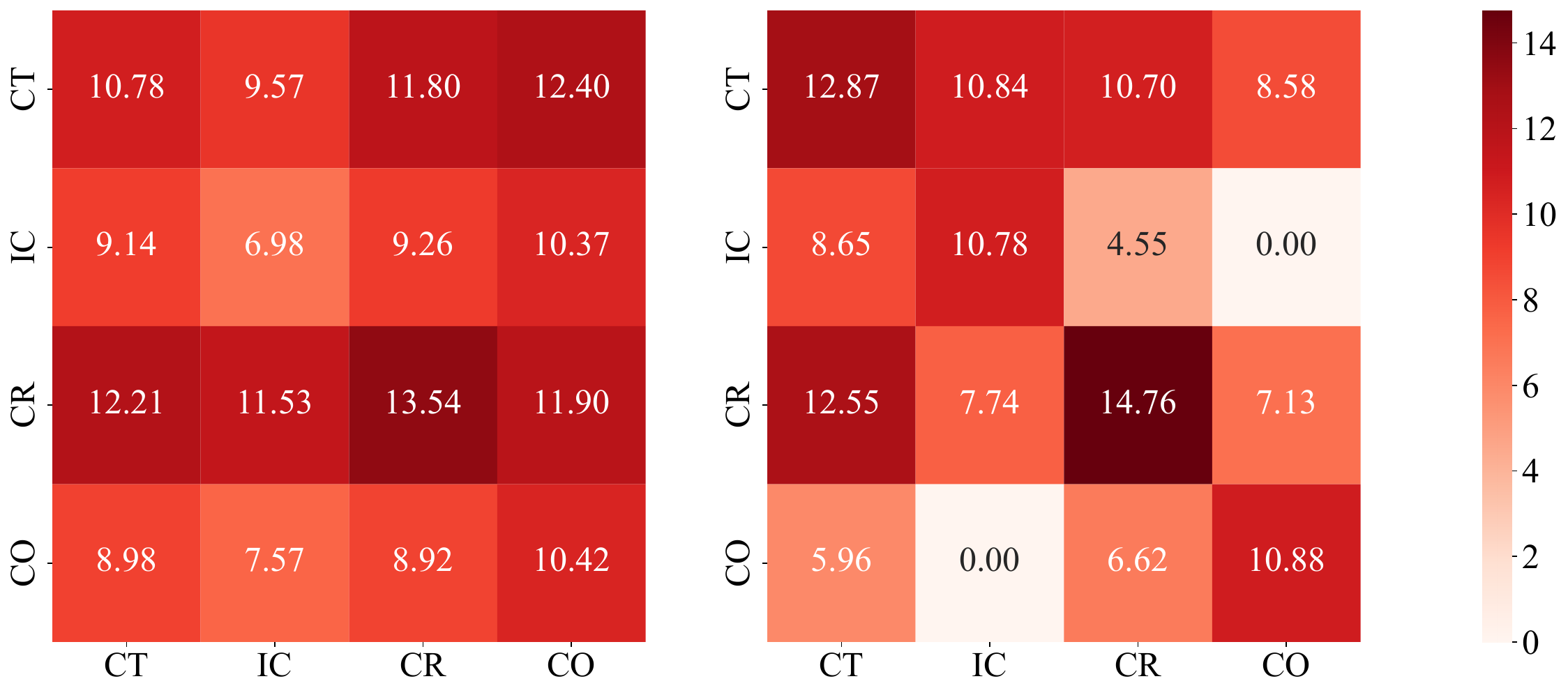}
\caption{\textbf{Comparison of confusion matrix for supervised and semi-supervised learning:} The image on the left shows the confusion matrix for supervised learning with 20\% labeled images and the right image shows semi-supervised learning with 20\% labeled images. The x-axis represents the predicted class labels and y axis represents the true class labels. The pixel values in the confusion matrices are transformed using natural logarithmic scale (base e) to enhance visibility and interpretation, especially due to the significant class imbalance in the dataset. The defects are abbreviated as CT: Contact, IC: Interconnect, CR: Crack and CO: Corrosion}

\label{fig:cm}
\end{figure}


We compare the supervised and semi-supervised learning models using a confusion matrix (Figure~\ref{fig:cm}), which uses a $log_{e}$ scale to enhance visual clarity. The supervised model misclassifies many pixels, resulting in low performance. In contrast, the semi-supervised model shows fewer misclassifications, leading to improved defect detection. The class imbalance, as shown in Figure~\ref{fig:piechart}, introduces some bias in the predictions. Specifically, the semi-supervised model correctly classifies a higher number of pixels for each defect class compared to the supervised model, demonstrating the effectiveness of the PV-S3 model. To provide quantitative insight, we report both the absolute number of correctly classified pixels and their natural logarithm (log\textsubscript{e}) values, as shown in the confusion matrix. For the semi-supervised model, the correctly classified pixel counts are: $12.87$ ($\log_{e}(387,157)$) for Contact, $10.78$ ($\log_{e}(48,175)$) for Interconnect, $14.76$ ($\log_{e}(2,582,157)$)) for Crack, and $10.88$ ($\log_{e}(53,188)$)) for Corrosion. In comparison, the supervised model correctly classifies $10.78$ ($\log_{e}(48,050)$) for Contact, $6.97$ ($\log_{e}(1,070)$) for Interconnect, $13.54$($\log_{e}(757,701)$) for Crack, and $10.42$ ($\log_{e}(33,429)$) for Corrosion. These numbers illustrate a clear improvement by the PV-S3 model, especially for low-frequency classes.

Despite being the most frequent defect in terms of both image count and pixel area (Figure~\ref{fig:piechart}), the semi-supervised model performs poorly on the \textit{Contact} class. This counterintuitive result stems from the physical interplay of degradation processes in PV modules. Contact failures, such as corrosion or front grid interruptions, often co-occur with other defects like \textit{Cracks} and \textit{Interconnect} failures due to shared stress and failure pathways. For instance, corrosion near solder pads can induce local stress that leads to cracking, while broken interconnects can cause localized heating that accelerates corrosion. These defects frequently overlap spatially and exhibit similar visual textures, complicating the model’s ability to distinguish among them—especially under limited supervision. The confusion matrix in Figure~\ref{fig:cm} highlights this challenge, showing that Contact pixels are often misclassified as Cracks. Even prior works using fully supervised models \cite{fioresi2021automated} trained on the same dataset achieve only 66\% precision and 51\% recall on the Contact class. This consistent difficulty across settings suggests that the issue is not merely due to label scarcity but is intrinsic to the complex, interacting nature of defects in real-world PV cells.

\paragraph{Practical Implications and Real-World Applications}
The proposed semi-supervised learning method using a mean teacher network demonstrates significant potential for real-world applications in the field of photovoltaics. By effectively detecting defects in photovoltaic (PV) cells using only 20\% of labeled data, our approach addresses a critical challenge in the industry: the scarcity and cost of obtaining fully annotated datasets. This reduction in the requirement for labeled data not only lowers the barrier for implementing advanced defect detection systems but also accelerates the deployment of quality control processes in PV manufacturing. The ability to accurately identify defects with limited labeled data can lead to more efficient production lines, early detection of faults, and overall improvement in the reliability and performance of PV systems. Furthermore, the robustness of our method to class imbalances and its adaptability to diverse datasets enhance its applicability across different types of PV cells and manufacturing conditions. By facilitating more accessible and cost-effective defect detection, our research has the potential to significantly impact the quality assurance processes in the photovoltaics industry, contributing to the advancement of renewable energy technologies.





\section{Conclusion, Limitations and Future Work}
\label{sec: Conclusion}

In conclusion, our research introduces PV-S3, a semi-supervised learning framework for defect detection in PV modules. PV-S3 effectively utilizes both labeled and unlabeled data, achieving accurate detection despite limited labeled samples.
The proposed approach demonstrates significant potential in automating defect detection, reducing manual efforts, and enhancing the reliability of PV systems. The proposed approach provides mean IoU of 67\%, precision of 72.50\%, recall of 78.90\%, and F1-score of 78.15\% on one of the largest defect detection dataset with merely 20\% labeled data, significantly outperforming existing state-of-the-art methods which use 100\% labeled data.

Our research also highlights certain limitations such as difficulty in boundary detection.
While PV-S3 provides effective guidance for defect classification, accurately delineating defect boundaries can be challenging due to the defect's complex and irregular nature. 


Looking ahead, future research should focus on enhancing detection accuracy and robustness which could lead to development of tools as used for other energy tasks \cite{VOYANT2018343,connolly2010review}. 
Additionally, efforts to collect diverse unlabeled data and employ active learning strategies could enhance the model's generalization capabilities. Furthermore, extending our approach to encompass a broader range of defect types in PV modules would render the system more comprehensive and adaptable to real-world scenarios. Such advancements could significantly contribute to the automation of defect detection, reducing manual labor, and ultimately enhancing the reliability of solar PV systems.
\section*{Declaration of competing interests}
\label{sec:dci}

The authors confirm that there are no existing financial conflicts or personal connections that could potentially be perceived as influencing the outcomes or interpretations presented in this paper.

\section*{Data Availability}
\label{sec:dci}

The datasets used in this study are publicly available. The UCF-EL dataset can be accessed at \url{https://github.com/ucf-photovoltaics/UCF-EL-Defect} and the CSB dataset can be accessed at \url{https://github.com/TheMakiran/BenchmarkELimages}. These datasets are freely available for research purposes.

\bibliography{ref.bib}
\bibliographystyle{plain}

\end{document}